\documentclass{emulateapj}

\usepackage{apjfonts}
\usepackage{graphicx}
\usepackage{epstopdf}

\slugcomment{To appear in the Astrophysical Journal, May 1 issue}

\shorttitle{Radiation pressure and virial black hole masses}
\shortauthors{Marconi et al.}

\newcommand{\wlLwlV}{\ensuremath{L_{5100}}}
\newcommand{\bolion}{\ensuremath{a}}
\newcommand{\bolV}{\ensuremath{b}}
\newcommand{\FWHB}{\ensuremath{V_{H\beta}}}
\newcommand{\LHB}{\ensuremath{L_{H\beta}}}
\newcommand{\fvest}{\ensuremath{\tilde{f}}}
\newcommand{\gvest}{\ensuremath{g}}
\newcommand{\1}{\ensuremath{^{-1}}}
\newcommand{\2}{\ensuremath{^{-2}}}
\newcommand{\SEC}{\ensuremath{\,\mathrm{s}}}
\newcommand{\ERG}{\ensuremath{\,\mathrm{erg}}}
\newcommand{\EV}{\ensuremath{\,\mathrm{eV}}}
\newcommand{\CM}{\ensuremath{\,\mathrm{cm}}}
\newcommand{\KM}{\ensuremath{\,\mathrm{km}}}
\newcommand{\K}{\ensuremath{\,\mathrm{K}}}
\newcommand{\Lo}{\ensuremath{\,\mathrm{L}_\odot}}
\newcommand{\Lsun}{\Lo}
\newcommand{\Mo}{\ensuremath{\,\mathrm{M}_\odot}}
\newcommand{\Msun}{\Mo}
\newcommand{\CHISQ}{\ensuremath{\chi^2}}
\newcommand{\MBH}{\ensuremath{M_\mathrm{BH}}}
\newcommand{\MBHO}{\ensuremath{M_\mathrm{BH,0}}}
\newcommand{\Lsph}{\ensuremath{L_\mathrm{sph}}}
\newcommand{\LEdd}{\ensuremath{L_\mathrm{Edd}}}
\newcommand{\wlLwl}{\ensuremath{\lambda L_\lambda}}
\newcommand{\LEddO}{\ensuremath{L_\mathrm{Edd,0}}}
\newcommand{\sig}{\ensuremath{\sigma}}
\newcommand{\sige}{\ensuremath{\sigma_\mathrm{e}}}
\newcommand{\RBLR}{\ensuremath{R_\mathrm{BLR}}}
\newcommand{\sigmaT}{\ensuremath{\sigma_\mathrm{T}}}
\newcommand{\Te}{\ensuremath{T_\mathrm{e}}}
\newcommand{\forb}[2]{\mbox{[#1\,\textsc{\lowercase{#2}}]}} 
\newcommand{\NH}{\ensuremath{N_\mathrm{H}}}
\newcommand{\ten}[1]{\ensuremath{10^{#1}}}
\newcommand{\xten}[1]{\ensuremath{\times 10^{#1}}}
\newcommand{\parfrac}[2]{\ensuremath{\left(\frac{#1}{#2}\right)}}
\newcommand{\HB}{\ensuremath{\mathrm{H}\beta}}
\newcommand{\OIII}{\forb{O}{III}}

\begin{document}

\title{The effect of radiation pressure on virial black hole mass estimates\\ and the case of Narrow Line Seyfert 1 galaxies}

\author{Alessandro Marconi\altaffilmark{1},
David J.~Axon\altaffilmark{2},
Roberto Maiolino\altaffilmark{3},
Tohru Nagao\altaffilmark{4},
Guia Pastorini\altaffilmark{1},
Paola Pietrini\altaffilmark{1},
Andrew Robinson\altaffilmark{2},
Guidetta Torricelli\altaffilmark{5}}

\altaffiltext{1}{Dipartimento di Astronomia e Scienza dello Spazio, Universit\'a degli Studi di Firenze, Largo E. Fermi 2, 50125 Firenze, Italy}
\altaffiltext{2}{Physics Department, Rochester Institute of Technology, 85 Lomb Memorial Drive, Rochester, New
York 14623, USA}
\altaffiltext{3}{INAF - Osservatorio Astronomico di Roma, Via Frascati 33, I-00040 Monte Porzio Catone, Italy}
\altaffiltext{4}{National Astronomical Observatory of Japan, 2-21-1 Osawa, Mitaka, Tokyo 181-8588, Japan}
\altaffiltext{5}{INAF-Osservatorio Astrofisico di Arcetri, Largo E. Fermi 5, 50125, Firenze, Italy.}

\begin{abstract}
We consider the effect of radiation pressure from ionizing photons on black hole (BH) mass estimates based on the application of the virial theorem to broad emission lines in AGN spectra. 
BH masses based only on the virial product $\Delta V^2R$ and neglecting 
the effect of radiation pressure can be severely underestimated especially in objects close to the Eddington limit.
We provide an empirical calibration of the correction for radiation pressure and we show that it is consistent with a simple physical model in which BLR clouds are optically thick to ionizing radiation and have average column densities of $\NH\sim \ten{23}\CM\2$. This value is remarkably similar to what is required in standard BLR photoionization models to explain observed spectra.
With the inclusion of radiation pressure the discrepancy between virial BH masses based on single epoch spectra and on reverberation mapping data drops from 0.4 to 0.2 dex rms. The use of single epoch observations as surrogates of reverberation mapping campaigns can thus provide more accurate BH masses than previously thought.
Finally, we show that Narrow Line Seyfert 1 (NLS1) galaxies have apparently low BH masses because they are radiating close to their Eddington limit. After the radiation pressure correction, NLS1 galaxies have BH masses similar to other broad line AGNs and follow the same \MBH-\sige/\Lsph\ relations as other active and normal galaxies. 
Radiation forces arising from ionizing photon momentum deposition constitute an important physical effect which must be taken into account when computing virial BH masses.
\end{abstract}


\defcitealias{vestergaard:2006}{V\&P06}
\defcitealias{onken:2004}{ONK+04}
\defcitealias{peterson:2004}{PET+04}
\section{Introduction}

In the last few years, it has become increasingly clear that supermassive black holes (BH) are an essential element in the evolution of galaxies. The key observational evidence of a link between a BH and its host galaxy is provided by the tight correlations between BH mass and luminosity, mass, velocity dispersion and surface brightness profile of the host spheroids (\citealt{kormendy:1995,gebhardt:2000,ferrarese:2000,marconi:2003b,graham:2007a}). The link between BH and host galaxy is probably established by the feedback of the accreting BH, i.e.~the active galactic nucleus, on the host galaxy itself (e.g.~\citealt{silk:1998,granato:2004,di-matteo:2005,croton:2006}, and references therein).

In order to fully understand the implications of BH growth on the evolution of the host galaxies it is fundamental to measure BH masses in large samples of galaxies from zero to high redshifts. Direct BH mass estimates based on stellar and gas kinematics are possible only in the local universe and their complexity does not allow their application to large samples (e.g.~\citealt{ferrarese:2005,marconi:2006}).
The limit of the local universe can be overcome with the reverberation mapping (RM) technique  (see, e.g., \citealt{peterson:2006} for a recent review) which provides an estimate of the Broad Line Region (BLR) average distance from the BH (\RBLR). The BH mass can thus be derived using the virial theorem,  $\MBH = f\Delta V^2\RBLR/G$, where $\Delta V$ is the width of the broad emission line and $f$ is a scaling factor which depends on the physical properties of the BLR (e.g.~\citealt{peterson:2000}).  Although this technique is potentially plagued by many unknown systematic errors (\citealt{krolik:2001};\citealt{collin:2006}), BH masses from reverberation mappping are in agreement with the \MBH-\sige\ relation of normal galaxies (e.g.~\citealt{mclure:2002}).
However, this technique is very demanding in terms of telescope time and it can be applied only to a few objects especially at high redshifts \citep{peterson:2004,kaspi:2007}.
The radius-luminosity relation discovered by \cite{kaspi:2000} shows that continuum luminosity can be used as a proxy for \RBLR\  
(\citealt{kaspi:2000,kaspi:2005,bentz:2006a}).
From the spectrum of a broad line AGN it is therefore possible to obtain a single epoch (SE) BH mass estimate.

One of the most important sources of uncertainty in virial \MBH\ estimates is the scaling factor $f$. \cite{onken:2004} have provided an estimate of $f$ assuming that the AGN in the RM database of \cite{peterson:2004} follow the \MBH-\sige\ relation of normal galaxies \citep{tremaine:2002,ferrarese:2005}. The factor $f$ by \cite{onken:2004} is only applicable to estimates of the virial product based on RM (see \citealt{peterson:2004} for more details). Building on the results by \cite{onken:2004}, \cite{vestergaard:2006} have calibrated scaling relations for SE virial \MBH\ estimates which combine the width of broad \HB\ with the luminosities of \wlLwl\ at 5100\AA.

Overall, SE virial estimates are commonly used to estimate BH masses in large sample of galaxies from zero to high redshifts (e.g.~\citealt{willott:2003,mclure:2004,vestergaard:2004,jiang:2007}) and are deemed accurate only from a statistical point of view on large samples of objects since a single measurement can be wrong even by a factor of  $\sim10$ (e.g.~\citealt{vestergaard:2006}).

There are three important considerations which are suggested by the results presented in the above papers.
First, SE virial BH masses of a few objects (e.g.~high $z$, high $L$ quasars or Narrow Line Seyfert 1 galaxies) imply they radiate near or above the Eddington limit. The virial theorem is based on the assumption that the system is gravitationally bound and this might be violated in super-Eddington sources where the outward force due to radiation pressure overcomes gravitational attraction.
Second, even when $L<\LEdd$, one should take into account that the radiation force partially compensates gravitational attraction. In the standard accretion disk model, the source of ionizing photons can be considered point-like at the distance of the BLR (see however \citealt{collin:2001} for a different point of view) and the radiation force scales as $r^{-2}$ mirroring the radial dependence of the BH gravitational attraction. Thus BLR clouds are effectively being pulled by a smaller effective BH mass and all present virial mass estimates for objects close to their Eddington limit, where radiation pressure is not considered, might be underestimated.
Finally, the Eddington limit is computed assuming that the radiation pressure is due only to Thomson scattering of photons by free electrons. As supported by reverberation mapping, by the radius-luminosity relation and other observational evidences (e.g.~\citealt{blandford:1990}), BLR clouds are almost certainly photoionized. Thus BLR clouds are subject to radiation forces arising from the deposition of momentum by ionizing photons which can substantially exceed that due to scattering.

The importance of radiation pressure due to ionizing photons and its possible effects on virial BH masses has already been mentioned in a few papers (e.g.~\citealt{mathews:1993,gaskell:1996}) but seem to have not been considered in detail subsequently.
This effect might be particularly important in Narrow Line Seyfert 1 galaxies which are believed to accrete close to their Eddington limit. Indeed, they are characterized by small BH masses compared to other AGNs and to the \MBH-\Lsph/\sige\ relations (e.g.~\citealt{mathur:2001}). It has also been noted that the distance of NLS1 galaxies from the \MBH-\Lsph/\sige\ relations is larger for objects with larger Eddington ratios (\citealt{grupe:2004}) suggesting that smaller BHs are growing faster. Alternatively, this might be an indication that that virial BH mass are underestimated in the high $L/\LEdd$ regime.

In this paper we investigate the effect of radiation pressure on virial BH mass estimates.
In \S\ \ref{sec:physics} we present a simple physical model for the radiation pressure effect on virial BH mass estimates.
In \S\ \ref{sec:observations} we calibrate the effect of radiation pressure on virial BH masses adapting the procedures of \cite{onken:2004} and \cite{vestergaard:2006}.
In \S\ \ref{sec:NLS1} we apply our corrected virial BH mass estimates to Narrow Line Seyfert 1 galaxies and show that these galaxies are indeed consistent with the \MBH-\sige/\Lsph\ relations, showing that BHs are not abnormally small.
Finally, we discuss our results and draw our conclusions in \S\ \ref{sec:summary}.

\section{The effect of radiation pressure on virial black hole mass estimates: a simple physical approach}\label{sec:physics}
%%%%%%% FIGURE 1%%%%%%%%%%%%%%%%%%%%%%%%%%%%%%%%%%%%%%%%%%
\begin{figure}
\centering
\includegraphics[angle=0, width=0.99\linewidth]{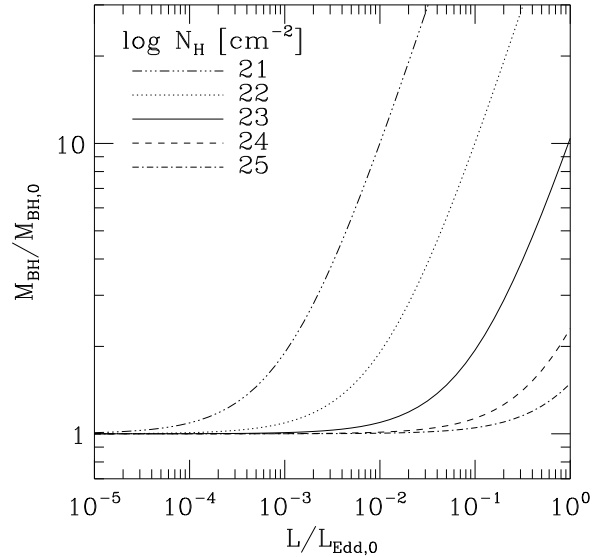}
%\plotone{f1.eps}
\caption{\label{fig:corrfac}The ratio between virial BH masses taking into account radiation pressure (\MBH) and those based only on the virial product (\MBHO) is plotted against the classical Eddington ratio based on \MBHO. \MBH/\MBHO\ is thus the correction factor which should be applied to BH mass estimates based only on the virial product. \NH\ is, on average, the total column density of each BLR cloud along the direction to the ionizing source.
}
\vspace{1cm}
\end{figure}
%%%%%%% FIGURE 1%%%%%%%%%%%%%%%%%%%%%%%%%%%%%%%%%%%%%%%%%%

We will explore the effect of radiation pressure on BLR clouds using a simplified model which assumes that (i) each cloud is optically thick to ionizing photons but optically thin to scattering processes, (ii) the Thomson cross-section is representative of all scattering processes involving free or bound electrons and (iii) both recombination and scattered photons are 'isotropically' re-emitted.
These assumptions are valid if $U c/\alpha_\mathrm{B}(H) < \NH < 1/\sigmaT$ where \NH\ is the total cloud column density along the direction to the ionizing source, $U$ is the ionization parameter, $\alpha_\mathrm{B}(H)$ is the 'case-B' recombination coefficient for hydrogen and \sigmaT\ is the Thomson cross-section. For typical conditions in the BLR ($\Te\simeq2\xten{4}\K$, $U\simeq 0.01$, e.g.~\citealt{netzer:2006}) $1.2\xten{21}\CM\2<\NH<1.5\xten{24}\CM\2$.
% alphaB at t=10000 K is 2.59e-13
% alphaB at t=20000 K is 2.52e-13

The total force acting on a cloud in the outward radial direction and due to radiation pressure is
\begin{equation}
F = \int_0^{+\infty}\mathrm{d}\nu\, \frac{L_\nu}{4\pi r^2\,c} (1-\mathrm{e}^{-\tau_\nu}) \Delta A 
\end{equation}

where $L_\nu$ is the luminosity of the AGN continuum emission,  $r$ is the cloud distance from the ionizing source, $\tau_\nu$ is the optical depth of absorption/scattering processes and $\Delta A$ is the cloud surface exposed to the AGN radiation. Scattering is important only for non-ionizing photons therefore, under the above assumptions, it is possible to write:
\begin{equation}
F = \frac{L_\mathrm{ion}}{4\pi r^2 c}\Delta A+\frac{L-L_{ion}}{4\pi r^2 c}\sigma_\mathrm{T} N_\mathrm{H}\Delta A
\end{equation}
where the two terms are the radiation forces due to absorption of ionizing photons and Thomson scattering, respectively, $L_\mathrm{ion}$ is the total luminosity of the AGN ionizing continuum, $h\nu>13.6\EV$ (see, e.g., \citealt{peterson:1997,krolik:1999}) and \NH\ is, on average, the total column density of each BLR cloud along the direction to the ionizing source. The contribution to the radiation force from the absorption of line photons is negligible for the optically thick clouds considered here (see, e.g., the seminal paper by \citealt{castor:1975}).  

Taking into account the total radiation force acting on each cloud and assuming that the BLR is a bound system, it is possible to derive a modified version of the classical virial theorem which takes into account radiation as well as gravitational forces.
Approximating the cloud mass as $\sim m_\mathrm{p} N_\mathrm{H}\Delta A$ the modified expression for the virial BH mass $M_\mathrm{BH}$ is:
\begin{equation}\label{eq:mvircorr}
\MBH = f \frac{V^2 r}{G}+\frac{L}{L_\mathrm{Edd,\odot}}\left[1-\bolion+\frac{\bolion}{\sigma_\mathrm{T}N_\mathrm{H}}\right]\Msun
\end{equation}
where $f$ is a geometrical factor which takes into account the geometry of the BLR, $L_\mathrm{Edd,\odot}$ is the classical Eddington luminosity for a solar mass object and 
$\bolion = L_\mathrm{ion}/L$. This expression has a physical meaning as long as the system is bound, i.e.~as long as the radiation force on BLR clouds is smaller than gravity. In formulae
\begin{equation}
L < \frac{\LEddO}{\left[1-\bolion+\bolion\,/(\sigma_\mathrm{T}N_\mathrm{H})\right]}
\end{equation}
where \LEddO\ is the classical Eddington luminosity.
Neglecting momentum injection by ionizing photons ($\bolion=0$) we recover the classical relation $L<\LEddO$.
Using \MBH\ from Eq.~\ref{eq:mvircorr} to compute \LEddO, it should be noticed that for $L\rightarrow\infty$, $L/\LEddO\rightarrow 1/\left[1-\bolion+\bolion\,/(\sigma_\mathrm{T}N_\mathrm{H})\right]$ and $L/\LEddO$ will always be less than or equal to 1. This is a consequence of the assumption of gravitationally bound BLR which allowed us to write Eq.~\ref{eq:mvircorr}. Therefore, it is not possible to establish whether a system is above Eddington by using virial BH mass estimates, since they are themselves based on the assumption of a sub-Eddington system. 

In order to quantify the effect of the radiation force correction we write Eq.~\ref{eq:mvircorr} as
\begin{equation}
\MBH =\MBHO\left[1+\frac{L}{\LEddO}\left(1-\bolion+\frac{\bolion}{\sigma_\mathrm{T}\NH}\right)\right]
\end{equation}
where \MBHO\ is the standard virial BH mass computed without taking into account radiation pressure.
In Fig.~\ref{fig:corrfac} we show the behavior of  \MBH/\MBHO\ as a function of  $L/ \LEddO$ and for different values of \NH.
The $a=L_\mathrm{ion}/L$ bolometric correction has been computed following \cite{marconi:2004} and is on average $\bolion\simeq 0.6$ in the $\ten{10}-\ten{12}\Lsun$ luminosity range.
For \NH=\ten{23}\CM\2\ and  $L/ \LEddO> 0.1$, \MBH/\MBHO\ varies between 2 and 10. This can be much larger for smaller column densities of BLR clouds but values at low \NH\ should be taken with caution since the adopted formula is valid only if the cloud is optically thick to ionizing photons, i.e.~$\NH>U c/\alpha_\mathrm{B}(H) \simeq 1.2\xten{21}(U/0.01)\CM\2$.
The correction factor remains small ($<2$) only for column densities $\NH>\ten{24}\CM\2$.
Clearly the correcting factor critically depends on the \NH\ value which sets the total cloud mass and thus the relative importance of gravitational attraction with respect to radiation pressure.
Overall, this figure suggest that neglecting the effect of radiation pressure might result in \MBH\ values which are underestimated even by a factor $\sim 10$.

Virial estimates of BH masses are based on the assumption that the BLR is gravitationally bound to the BH and that outflowing motions are negligible.  In recent years, building upon observational evidence for outflows in the BLR, alternative models have been proposed in which part of the BLR is in the form of a disk wind (e.g.~\citealt{murray:1995a,chiang:1996,elvis:2000,collin:2001,proga:2000,proga:2007b,everett:2005} and references therein). This possibility has generated a debate about the reliability of virial BH masses (e.g.~\citealt{peterson:2000,krolik:2001,onken:2002,collin:2006,vestergaard:2006}) which is beyond the scope of this paper. 
Nevertheless,  virial BH mass estimators are widely used, and  
in order to investigate the effect of radiation pressure on such estimates, we must necessarily start from the same set of assumptions for our  simple model.

\section{The effect of radiation pressure on virial black hole mass estimates: an observational approach}\label{sec:observations}

The simple physical approach presented in the previous sections suggests that  virial BH mass estimates can be written as a function of observed quantities as
\begin{equation}\label{eq:new}
\MBH = f\, \frac{V^2 R}{G}+g\,\left(\frac{\wlLwlV}{\ten{44}\ERG\SEC\1}\right)\Msun
\end{equation}
where \wlLwlV\ represents \wlLwl\ at 5100\AA. After Eq.~\ref{eq:mvircorr}, $g$ corresponds to
\begin{equation}\label{eq:gfactor}
g = 6.0\xten{6}\,\parfrac{\bolV}{9.0}\left(1-\bolion+\frac{\bolion}{\sigma_T\NH}\right)
\end{equation}
where $\bolV = L/\wlLwlV$ is the bolometric correction at 5100\AA.
Following \cite{marconi:2004}, the $L/\wlLwlV$
bolometric correction is on average $\bolV\simeq 9.0$ in the $\ten{10}-\ten{12}\Lsun$ luminosity range.
$f$ and $g$ are free unknown parameters which depend on the physical and geometrical properties of the BLR.
In particular the $g$ factor critically depends on the assumed \NH\ value which determines the cloud mass and thus sets the relative importance of gravity and radiation pressure.

A correction for the radiation force which is proportional to $L$ is more general than the simple physical model presented in the previous section therefore, in order to avoid \textit{a-priori} assumptions on the values of the physical parameters characterizing BLR clouds, we can determine $f$ and $g$ following a procedure similar to \cite{onken:2004} and \cite{vestergaard:2006}. Thus our model will only provide a simple physical interpretation of the empirical $g$ values.

\subsection{Black hole masses from reverberation mapping data}
\cite{onken:2004} considered the AGNs from the reverberation mapping database by \cite{peterson:2004} with measured stellar velocity dispersion. They used the time lag of the broad lines for $R$ and the velocity dispersion of the $r.m.s.$ spectra for $V$. They determined $f$ by assuming that the AGNs in their sample follow the \MBH-\sige\ relation for normal galaxies.

We first update the RM database by \cite{peterson:2004} with the newer estimates of BLR time lags for NGC 4151 \citep{bentz:2006}, NGC  4593 \citep{denney:2006} and NGC5548 \citep{bentz:2007}. We exclude from the database PG1211+143 and IC4329A because their time lags are not reliable (\citealt{peterson:2004}).
When possible, we correct the average AGN luminosities \wlLwl(5100 \AA) for the host galaxy contamination following \cite{bentz:2006a}.
We consider the host galaxy velocity dispersions by  \cite{onken:2004} and we supplement them with the data by \cite{nelson:2004} for Mrk 279 and \cite{dasyra:2007} for PG1229+204, PG1426+015, PG1617+175 and PG2130+099.

$f$ and $g$ are then derived by finding the minimum of:
\begin{equation}\label{eq:chisq}
\CHISQ = \sum_i \frac{[\,(\log\MBH)_i-(\log\MBH)_{0,i}]^2\,}{(\delta \log\MBH)^2_i+(\delta \log\MBH)^2_{0,i}+\Delta\Sigma^2}
\end{equation}
where $(\log\MBH)_i$ is the log BH mass of the $i$-th object which depends on $f$ and $g$,
$(\log\MBH)_{0,i}=\alpha+\beta\log(\sige/200\KM\SEC\1)_i$ is the expected mass value from the \MBH-\sig\ relation \citep{tremaine:2002,ferrarese:2005}.
 $\sige$ is the stellar velocity dispersion of the host spheroid, $(\delta\log \MBH)_i$ is the error on $(\log\MBH)_i$ based on the errors on $V^2$, $R$ and $\delta(\log\sige)_i$ is the error on $\log(\sige/200)_i$.
At variance with  \cite{onken:2004}, we allow for an intrinsic dispersion of the \MBH-\sig\ relation, $\Delta\Sigma$, which we  assume equal to 0.25 dex (e.g.~\citealt{tremaine:2002,marconi:2004,tundo:2007}).
We follow a standard \CHISQ\ minimization and we estimate errors on the parameters with the bootstrap method \citep{efron:1994} with 1000 realizations of the parent sample.
As shown by \cite{onken:2004} the use of the \cite{ferrarese:2005} or \cite{tremaine:2002} version of the \MBH-\sige\ relation provides consistent results; therefore, in the following we will focus only on the \cite{tremaine:2002} relation, $\alpha=8.13\pm0.06$, $\beta=4.02\pm0.32$.
%%%% TABLE1%%%%%%%%%%%%%%%%%%%%%%%%%%%%%%%%%%%%%%%%%%
\begin{deluxetable}{lcccr} 
\tablecolumns{4}
\tablewidth{0pc}
\tablecaption{Calibration of RM virial masses\label{fit:onken}}
\tablehead{
Database &  \colhead{$f$} & \colhead{$\log g$} & $\Delta_\mathrm{res}$  }
\startdata
\\
Onken2004\tablenotemark{\ensuremath{\dagger}}		& $5.5\,\,(+1.9;-1.5)$	& $-10\tablenotemark{\ensuremath{\star}}$	&	$0.39$	\\
Onken2004 								& $5.2\,\,(+1.6;-1.2)$	& $-10\tablenotemark{\ensuremath{\star}}$	&	$0.39$	\\
Updated									& $4.8\,\,(+1.5;-1.3)$ 	& $-10\tablenotemark{\ensuremath{\star}}$	&     	$0.52$	\\
Updated \hfill	(Fam1)						& $\mathbf{3.1\,\,(+1.3;-1.5)}$ & $\mathbf{7.6\,\,(+0.3;-0.3)}$	&     	$0.50$	\\
	      \hfill	(Fam2)     					& $4.3\,\,(+1.2;-1.1)$ & $< 2$	&     		\\
\enddata
\tablenotetext{\ensuremath{\dagger}}{$\Delta\Sigma=0.0$ as in \cite{onken:2004}.}
\tablenotetext{\ensuremath{\star}}{Fixed fit parameter.}
\end{deluxetable}
%%%% TABLE1%%%%%%%%%%%%%%%%%%%%%%%%%%%%%%%%%%%%%%%%%%

The results of the fitting procedure are summarized in Table \ref{fit:onken}. We have considered the original \cite{onken:2004} database and the updated one.
Errors on fit parameters are determined from the percentiles of the bootstrap results at the 68\% confidence level around the median.
Several considerations can be made from the results in Table \ref{fit:onken}.
As a sanity check, we are able to reproduce the results by \cite{onken:2004} i.e.~$f=5.5\pm 1.9$ (first table row).
The fits shown in the second and third row indicate that when $g$ is fixed and negligible, the use of the updated database 
or the use of an intrinsic dispersion for \MBH-\sig\ do not significantly change the $f$ value. With the use of the updated database which has a larger number of objects, the scatter of the residuals is significantly increased.
When $g$ is free to vary, the bootstrap analysis shows that there are two distinct families of solutions: those where both $f$ and $g$ are determined and those where $g$ is negligible and totally undetermined.
The existence of two families of solutions from the bootstrap simulations is an indication that the dependence on luminosity can be inferred only from part of the sample, i.e.~from the objects with the largest $L/\LEdd$ ratios. In roughly 20\%\ of the sample realizations the number of these objects is low, $g$ is undetermined and the $f$ values are consistent with the \cite{onken:2004} determination.
The inclusion of the $g$ parameter has the net effect of decreasing $f$, since the expected BH mass is fixed by the \MBH-\sige\ relation.

Our ability to determine an accurate empirical value of $g$ is limited, as were previous efforts to determine $f$, by the size, composition and accuracy of the existing reverberation database. In particular, it currently contains few sources with high Eddington ratios, which provide the tightest constraints on $g$. With this caveat in mind, however, we provide a first estimate of $f=3.1\pm1.4$ and $\log g=7.6\pm0.3$ to compute \MBH\ from reverberation mapping data.

\subsection{Black Hole masses from single epoch spectra}
%%%%%%% TABLE 2%%%%%%%%%%%%%%%%%%%%%%%%%%%%%%%%%%%%%%%%%%
\begin{deluxetable}{lcccr}
\tablecolumns{4}
\tablewidth{0pc}
\tablecaption{Calibration of SE virial masses\label{fit:vestergaard}}
\tablehead{\MBH\ from RM  & \colhead{$\log\fvest$} & \colhead{$\log\gvest$} & $\Delta_\mathrm{res}$  }
\startdata
\cutinhead{\MBH\ from SE (\FWHB, \wlLwlV)}
$f=5.5;\, \log g=-10$\tablenotemark{\ensuremath{\dagger}}     &   $6.93\,\,(+0.12;-0.13)$       &       $-10\tablenotemark{\ensuremath{\star}}$      &   $0.43$    \\ 
$f=5.5;\, \log g=-10$     &   $6.47\,\,(+0.17;-0.22)$       &       $7.48\,\,(+0.16;-0.25)$      &   $0.34$    \\ 
$f=3.1;\, \log g=7.6$     &   $\mathbf{6.13\,\,(+0.15;-0.30)}$       &       $\mathbf{7.72\,\,(+0.06;-0.05)}$      &   $0.22$    \\ 
\cutinhead{\MBH\ from SE (\FWHB, \LHB)}
$f=5.5;\, \log g=-10$\tablenotemark{\ensuremath{\dagger}}     &   $6.69\,\,(+0.12;-0.08)$       		&       $-10\tablenotemark{\ensuremath{\star}}$      	     &   $0.46$    \\ 
$f=3.1;\, \log g=7.6$     &   $\mathbf{5.95\,\,(+0.12;-0.17)}$      &       $\mathbf{7.82\,\,(+0.07;-0.09)}$      &   $0.27$    \\
\enddata
\tablenotetext{\ensuremath{\dagger}}{Original \cite{onken:2004} database.}
\tablenotetext{\ensuremath{\star}}{Fixed fit parameter.}
\end{deluxetable}
%%%%%%% TABLE 2%%%%%%%%%%%%%%%%%%%%%%%%%%%%%%%%%%%%%%%%%%
\cite{vestergaard:2006} considered the AGNs in the \cite{peterson:2004} database. They collected single epoch spectra for the same sources and used the FWHM of the broad lines as an estimate of $V$ and the continuum or broad line luminosity to estimate $R$ from the radius-luminosity relation of \cite{bentz:2006a}. Then they determined the corresponding $\fvest$ parameter (see Eq.~\ref{eq:SEnew} below) by rescaling the virial products from single epoch spectra to the BH masses determined following \cite{onken:2004}.
%%%%%%% FIGURE 2%%%%%%%%%%%%%%%%%%%%%%%%%%%%%%%%%%%%%%%%%%
\begin{figure*}
\centering
\includegraphics[angle=0, width=0.45\textwidth]{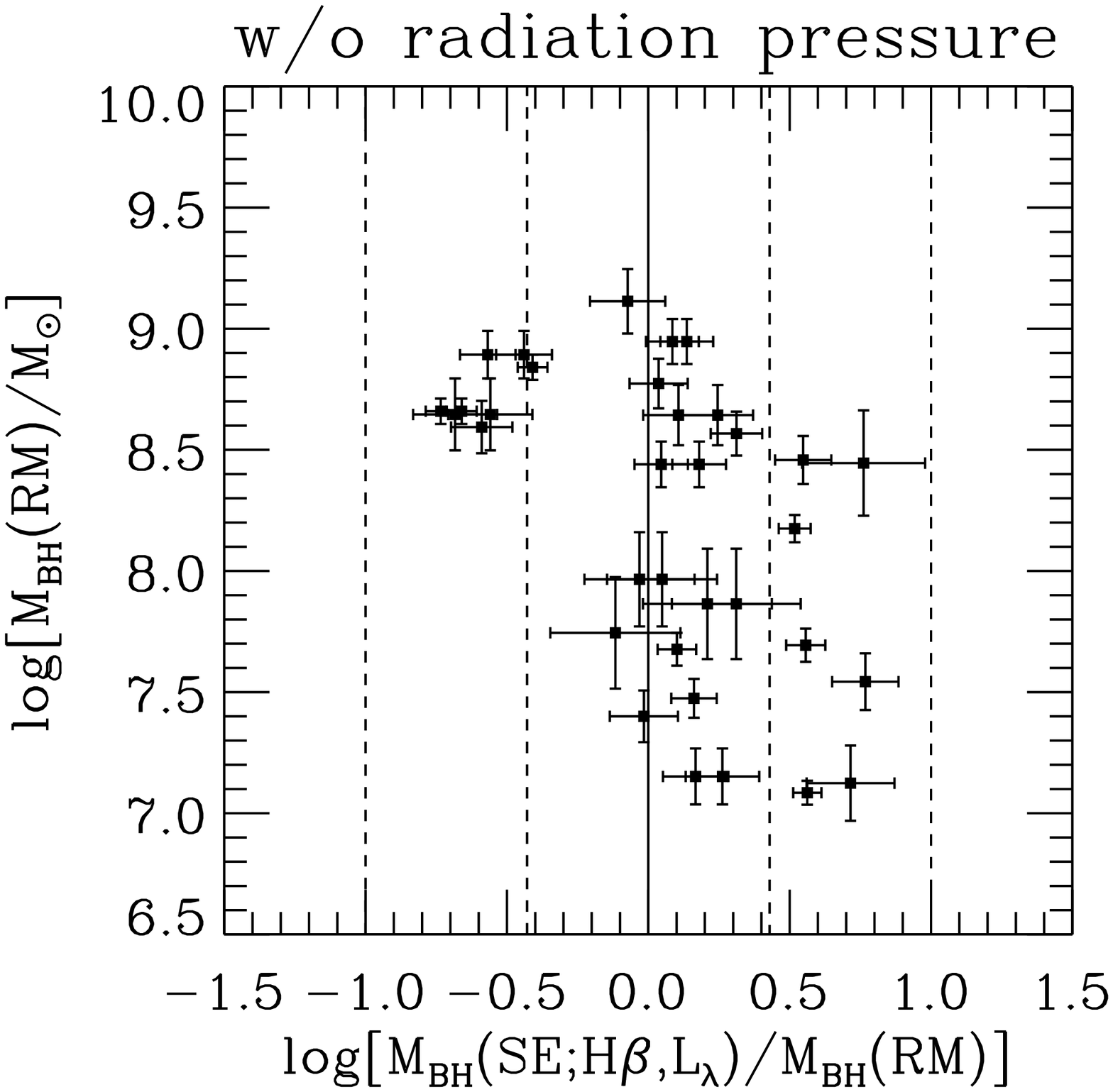}
\includegraphics[angle=0, width=0.45\textwidth]{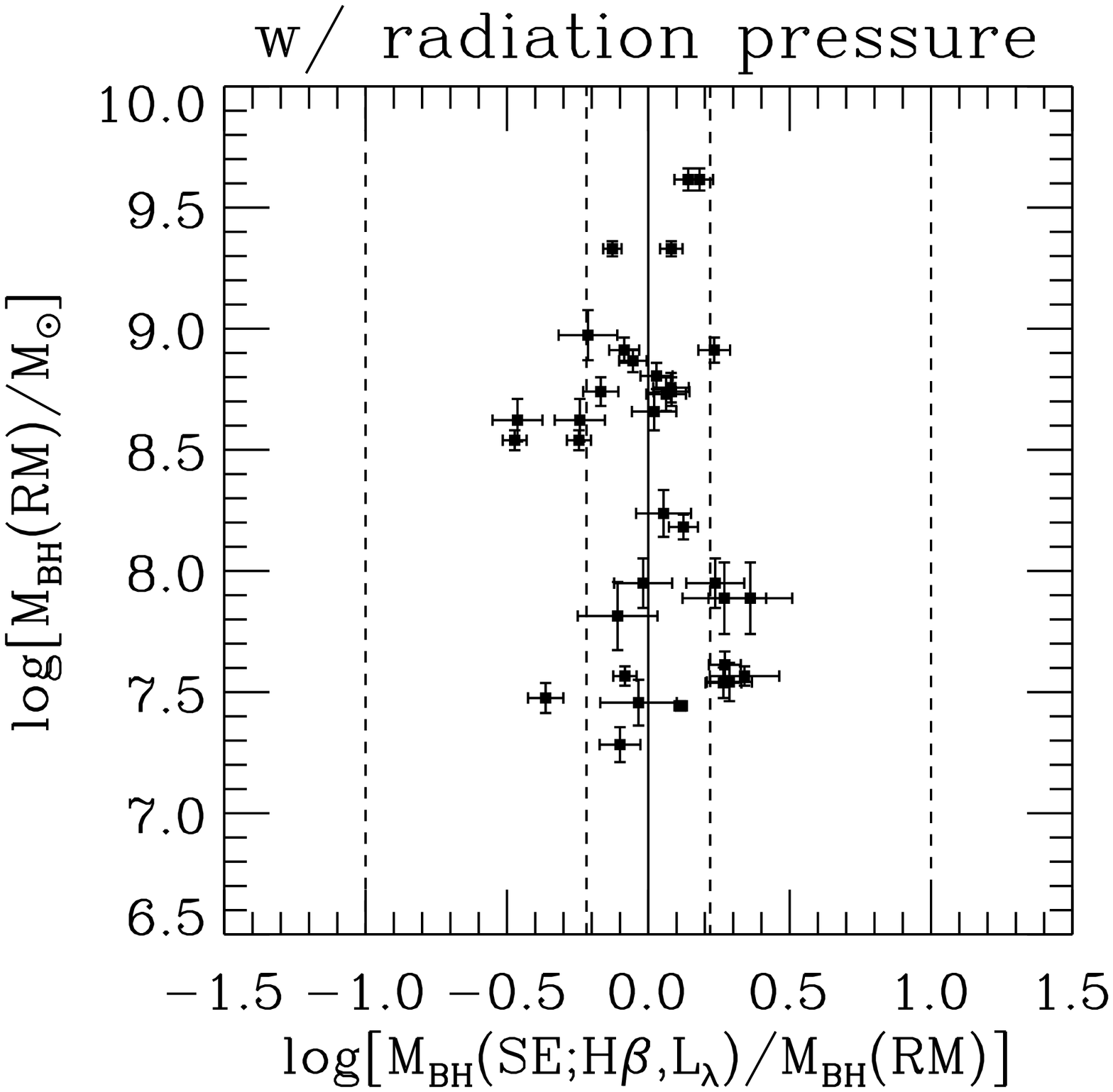}
%\plottwo{f2a.eps}{f2b.eps}
\caption{\label{fig:vester} Comparison between SE and corresponding RM virial masses. Left panel: without taking into account radiation pressure (same as Fig.~8, right panel, of \citealt{vestergaard:2006}); the dispersion of the data along the x axis around the 0 value is 0.4 dex. Right panel: same as right panel but taking into account radiation pressure as described in the text (RM virial masses are also computed with the correction for radiation pressure); the dispersion is 0.2 dex. 
Error bars combine errors on RM and SE virial masses but are dominated by the former. Error bars are different in the left and right panels because of the different relative importance of virial products and luminosities in RM virial masses.  
}
\end{figure*}
%%%%%%% FIGURE 2%%%%%%%%%%%%%%%%%%%%%%%%%%%%%%%%%%%%%%%%%%

We consider the database of single epoch measurements of $FWHM(\HB)$ (hereafter \FWHB), \LHB\ and \wlLwlV\ by \cite{vestergaard:2006} and following those authors we write the virial BH mass from single epoch measurements as: 
\begin{eqnarray}\label{eq:SEnew}
\MBH/\Msun &=& 
\fvest\, \parfrac{\FWHB}{1000\KM\SEC\1}^2\left(\frac{\wlLwlV}{\ten{44}\ERG\SEC\1}\right)^{0.5}\nonumber\\
 & & +\gvest\,\left(\frac{\wlLwlV}{\ten{44}\ERG\SEC\1}\right)
\end{eqnarray}
where the proxy for $V$ is now the FWHM of the \HB\ line and the BLR radius $R$ is given by the radius-luminosity relation with a slope of $0.50\pm 0.06$.
As before, the best $\fvest$ and $\gvest$ values follow from \CHISQ\ minimization as in Eq.~\ref{eq:chisq} where $(\log\MBH)_{0,i}$ is now the BH mass from reverberation mapping computed according to \citealt{onken:2004} ($f=5.5$, $\log g=-10.0$) or to our new calibration ($f=3.1$, $\log g=7.6$). Obviously, the $\Delta\Sigma$ term has been removed.

The fit results are shown in Table \ref{fit:vestergaard} where, as before, we provide bootstrap errors.
The fit results in the first row are the sanity check to show that we are able to reproduce the results by \cite{vestergaard:2006}, who find $\log \fvest = 6.91\pm 0.02$ with an rms of 0.43. 
Our errors are larger because of bootstrap simulations, but they would be similar to the ones by \cite{vestergaard:2006} if we used the formal errors of the fit.
In the second row we start from the assumption that virial masses from RM are computed following \cite{onken:2004}, but we allow for a free $\gvest$ factor. The SE data are clearly able to provide an estimate of the $\gvest$ factor which turns out to be remarkably similar to what was found for the RM data.
In the third row we start from virial RM masses computed with the best $f$ and $g$ values and there are two surprising results: first, the $\gvest$ value which turns out for SE virial masses is $\log \gvest = 7.72\pm 0.05$, perfectly consistent with that from RM virial masses, but with a much smaller uncertainty. Second, the dispersion of the residuals drops from $\sim 0.4$ to 0.2 dex.
The latter result indicates that half of the scatter of SE virial BH masses around RM ones is consistent with a need to take into account radiation pressure. 
The reduced scatter of the SE virial masses is also shown in Fig.~\ref{fig:vester} (right panel) and should be compared with the left panel in the same figure and Fig.~8 (right panel) of \cite{vestergaard:2006}.

\cite{wu:2004} and \cite{greene:2005} have shown that it is also possible to use the luminosity of the broad \HB\ instead of \wlLwlV\ to avoid possible contamination of the AGN continuum emission from the host galaxy.
Thus, following \cite{vestergaard:2006}, we substitute $\wlLwlV$ with \LHB\ in Eq.~\ref{eq:SEnew} to obtain SE virial masses from the broad \HB\ line only.
Inverting the $\LHB-\wlLwlV$ relation by \cite{greene:2005} we can write
\begin{eqnarray}\label{eq:SEnew2}
\MBH/\Msun & = & \fvest\, \parfrac{\FWHB}{1000\KM\SEC\1}^2\parfrac{\LHB}{\ten{42}\ERG\SEC\1}^{0.44} \nonumber\\
& & +\gvest\,\,0.732\parfrac{\LHB}{\ten{42}\ERG\SEC\1}^{0.883}
\end{eqnarray}
The fit results are shown in Table \ref{fit:vestergaard}. As before, we can reproduce the \cite{vestergaard:2006} calibration, $\log\fvest=6.67\pm 0.03$,  and the best fit which takes into account radiation pressure shows a significant drop in the dispersion of the residuals providing a best fit $\gvest$ value which is consistent with previous results.

\subsection{The average column density of BLR clouds}

The results in the previous sections show that it is possible to determine $f$ and $g$ both for RM and SE virial masses although it is difficult to accurately quantify their magnitude with the present data.
The $f$ values are smaller than those derived by \cite{onken:2004} and \cite{vestergaard:2006} because the final BH masses are still calibrated with the \MBH-\sig\ relation but part of the final \MBH\ value is accounted for by the effect of radiation pressure.
Considering the effect of radiation pressure can significantly improve the agreement of SE and RM virial masses.

The two $g$ values determined by (1) minimising the RM virial mass against the $\MBH-\sige$ relation and (2) minimising the SE virial mass against the 'calibrated' RM mass are both consistent with a value $\log g \simeq 7.7$.
Considering Eq.~\ref{eq:gfactor} we can derive the average \NH\ which is needed to obtain the $g$ value determined empirically.
With $\log g=7.7$ and $\bolion=0.6$ we can derive $\NH\simeq 1.1\xten{23}\CM\2$. This \NH\ value which we inferred by calibrating RM and SE virial BH masses is remarkably similar with the indications from photoionization modeling studies of the BLR.
Within the framework of the standard BLR model, photoionization calculations can explain observed spectra only if BLR clouds are optically thick to ionizing radiation and adopted \NH\ are usually of the order of \ten{23}\CM\2\ (e.g.~\citealt{baldwin:1995,kaspi:1999,korista:2004} and references therein).

\section{The case of Narrow Line Seyfert 1 galaxies}\label{sec:NLS1}
%%%%%%% FIGURE 3%%%%%%%%%%%%%%%%%%%%%%%%%%%%%%%%%%%%%%%%%%
\begin{figure*}
\centering
\includegraphics[width=0.95\textwidth]{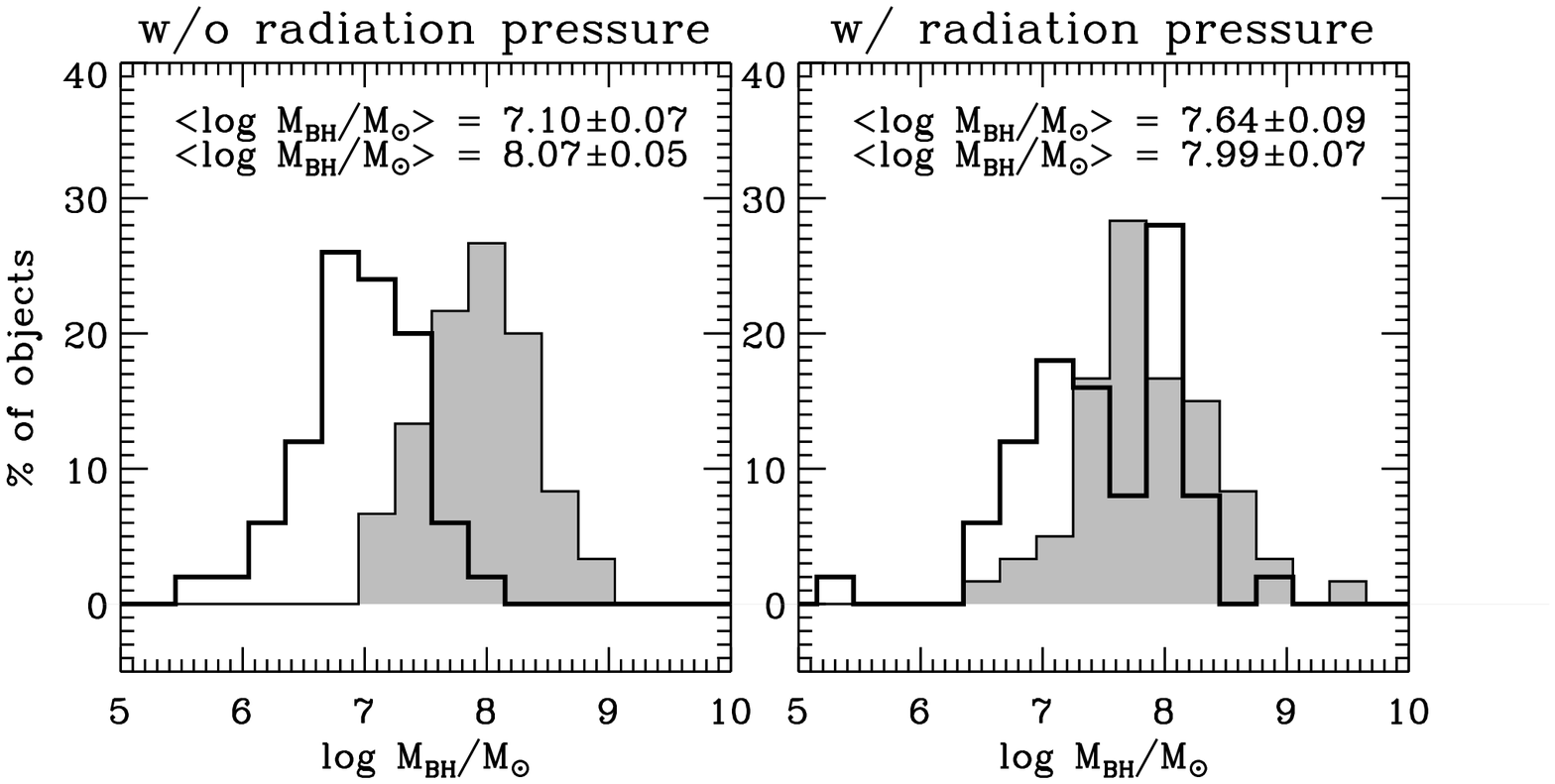}
%\plotone{f3.eps}
\caption{\label{fig:bhmass} Distributions of BH masses for narrow line Seyfert 1 galaxies ($\FWHB\le 2000\KM\SEC\1$, thick line) and 'normal' Seyfert 1 galaxies ($\FWHB>2000\KM\SEC\1$, thin line with shaded area). Left panel:  \MBH\ obtained with the scaling relations by \cite{vestergaard:2006}.
Right panel: \MBH\ obtained with the scaling relations which take into account radiation pressure.
The numbers in the top left corners of both panels denote the mean and standard deviation of the mean ($\sigma/\sqrt{N}$) of narrow and broad Seyfert 1 galaxies. }
\end{figure*}
%%%%%%% FIGURE 3%%%%%%%%%%%%%%%%%%%%%%%%%%%%%%%%%%%%%%%%%%

The nature of Narrow Line Seyfert 1 galaxies and their relation to 'normal' Seyfert 1 galaxies is still debated, but it is more or less generally believed that they are AGNs characterized by high accretion rates and small BH masses accounting for their smaller line widths (e.g.~\citealt{pounds:1995}).
Many different authors have undertaken the task of measuring virial BH masses in NLS1 galaxies and found that they are small compared to broad line AGNs with similar luminosities (e.g.~\citealt{grupe:2004a}). The location of NLS1 on the \MBH-\sige/\Lsph, however, is still hotly debated. Most authors suggest that NLS1 galaxies have small BHs compared to their host galaxies (e.g.~\citealt{mathur:2001,grupe:2004,zhou:2006,ryan:2007}) while others find an overall agreement with the \MBH-\sige\ relation of normal galaxies (e.g.~\citealt{botte:2005,komossa:2007}). 
A picture is now emerging in which the BHs in NLS1 galaxies are now experiencing a rapid growth which will eventually lead them on the \MBH-\Lsph/\sige\ relations as other active and normal galaxies (e.g.~\citealt{collin:2004,mathur:2005}).

NLS1 galaxies are thus ideal targets to explore the effects of the newly calibrated expressions which take into account radiation pressure.
In particular, using our new calibrated expressions for virial BH masses, we will verify (i) whether BH masses of NLS1 galaxies are indeed small compared to other AGNs with similar luminosities and (ii) whether they lie below the \MBH-\sige,\Lsph\ relations.

We first test whether BH masses in NLS1 galaxies are on average smaller than those in 'normal' Seyfert 1 galaxies. 
We consider the complete, soft X-ray selected sample by \cite{grupe:2004b} which is composed of 110 broad line AGNs with measured \FWHB, \LHB\ and \wlLwlV\ and we compute virial BH masses using Eq.~\ref{eq:SEnew}.
In Fig.~\ref{fig:bhmass} we plot the distributions of \MBH\ obtained with the scaling relations by \cite{vestergaard:2006} (left panel) and with the scaling relations which take into account radiation pressure (right panel).
The sample has been divided in two parts, narrow line Seyfert 1 galaxies ($\FWHB\le 2000\KM\SEC\1$, thick line) and 'normal' Seyfert 1 galaxies ($\FWHB>2000\KM\SEC\1$, thin line with shaded area).
In the top left corners of both panels we report the mean and standard deviation of the mean ($\sigma/\sqrt{N}$) of narrow and broad Seyfert 1 galaxies. If radiation pressure is not taken into account, we recover the well known result that BH masses are a factor $\sim 10$ smaller in NLS1 galaxies. However, this difference is greatly reduced to a factor $\sim 2$ when radiation pressure is taken into account.
The average BH mass of 'normal' Seyfert 1 galaxies is unchanged as expected since these objects are accreting at moderately low Eddington ratios compared to NLS1. It is beyond the scope of this paper to accurately determine the average BH mass of NLS1 with respect to Seyfert 1 galaxies, we only wish to point out that the effect of radiation pressure is very important and, when taken into account, BH masses of NLS1 galaxies are, on average, a factor 5 larger.  
%%%%%%% FIGURE 4%%%%%%%%%%%%%%%%%%%%%%%%%%%%%%%%%%%%%%%%%%
\begin{figure*}
\centering
\includegraphics[width=0.95\textwidth]{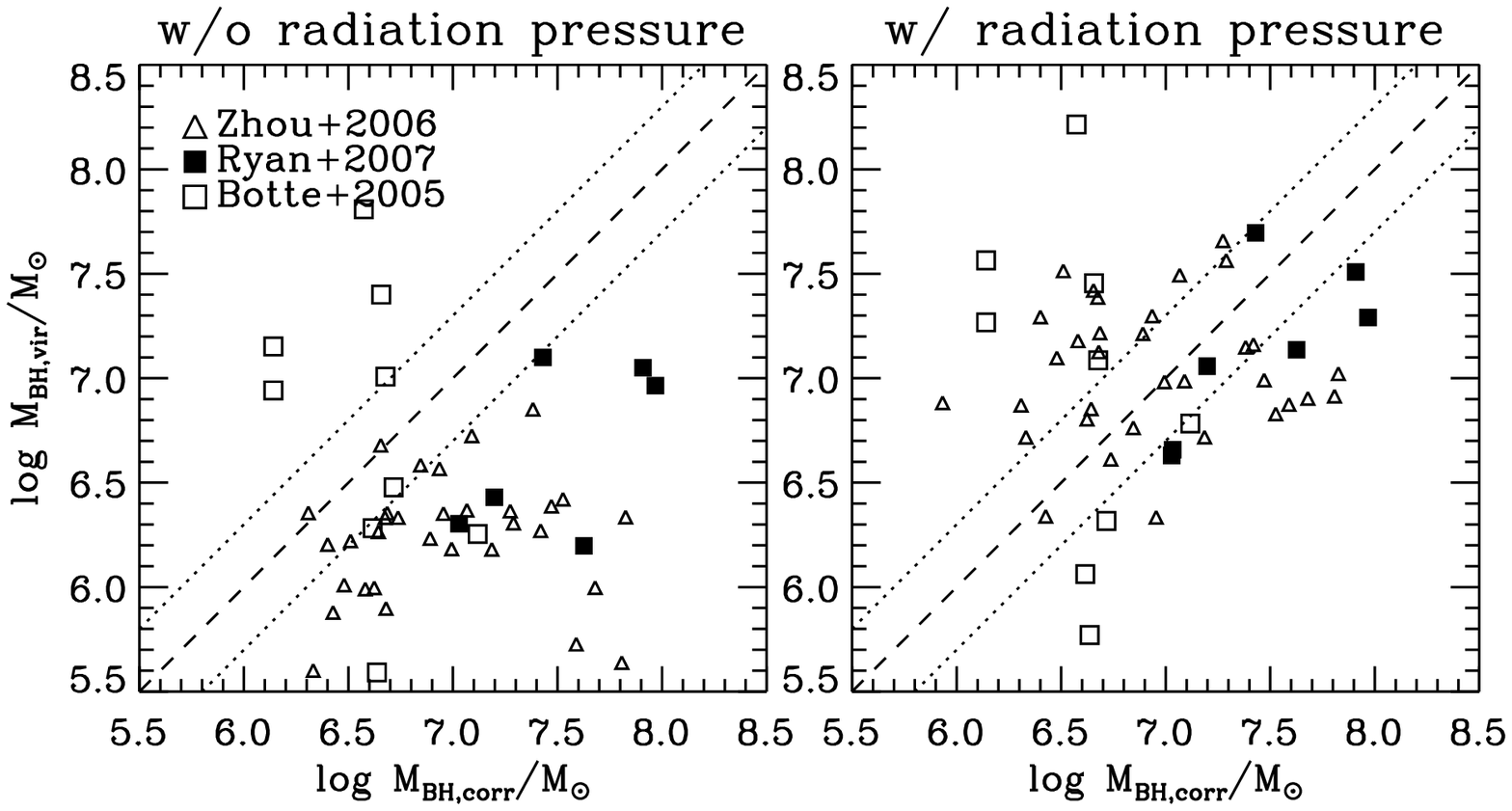}
%\plotone{f4.eps}
\caption{\label{fig:correlations} Comparison between virial BH masses ($M_\mathrm{BH,vir}$)
and those expected from the \MBH-\sige\ or \MBH-\Lsph\ correlations for normal galaxies ($M_\mathrm{BH,corr}$, \citealt{tremaine:2002,marconi:2003b}). Left panel: virial BH masses are computed using the calibrated relations by \cite{vestergaard:2006}. Right panel: virial BH masses are computed using the relations derived in this paper which take into account radiation pressure.}
\end{figure*}
%%%%%%% FIGURE 4%%%%%%%%%%%%%%%%%%%%%%%%%%%%%%%%%%%%%%%%%%

We now test whether NLS1 galaxies indeed lie below the \MBH-\sige/\Lsph\ relations.
We consider only samples where \sige\ or \Lsph\ are measured directly because we want to avoid issues connected with using \sige\ surrogates like the dispersion of the \OIII\ line (e.g.~\citealt{greene:2005a,komossa:2007}).
We thus consider the samples of NLS1 galaxies by \cite{botte:2005} and \cite{zhou:2006} where \sige\ are directly measured and the sample by \cite{ryan:2007}, the only one for which accurate high resolution J and K photometry of the host spheroid is available.
From \cite{zhou:2006} we take the sub-sample of 33 sources with z <0.1 for which 
either the host galaxy appears to be face on or the SDSS fiber 
aperture is dominated by galactic bulge contribution. This choice is motivated by the need to avoid bulge velocity dispersion values which are artificially increased by rotation of the galactic disks.

\begin{figure}
\centering
\includegraphics[width=0.95\linewidth]{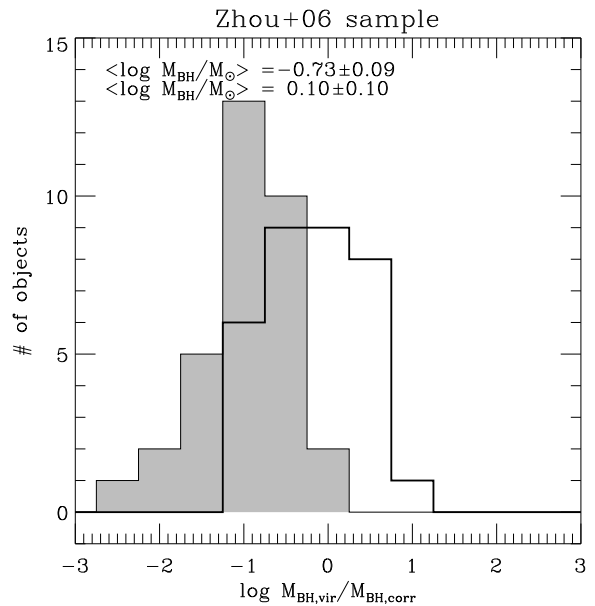}
%\plotone{f5.eps}
\caption{\label{fig:histocorr} Histogram of the $M_\mathrm{BH,vir}/M_\mathrm{BH,corr}$ ratio for the \cite{zhou:2006} sample of 33 NLS1 galaxies (see text).  Virial BH masses are computed using the calibrated relations by \cite{vestergaard:2006} (thin line with shaded area) and with the relations derived in this paper which take into account radiation pressure.}
\end{figure}

For the  \cite{botte:2005} and  \cite{ryan:2007} samples, we compute virial BH masses using  the scaling relations by \cite{vestergaard:2006} and Eq.~\ref{eq:SEnew}. Instead for the \cite{zhou:2006} sample we use Eq.~\ref{eq:SEnew2}, i.e.~we use the luminosity of the broad \HB\ as a proxy for \RBLR, 
since, due the latter selection criteria, \wlLwl\ might be strongly contaminated by stellar light.
The comparison with expected BH mass values from the \MBH-\sige\ \citep{tremaine:2002} and \MBH-\Lsph\ \citep{marconi:2003b} are plotted in the Fig.~\ref{fig:correlations}: in the left panel we use the virial BH masses by \cite{vestergaard:2006} while in the right panel we use our new virial mass estimates which take into account radiation pressure.
A more refined statistical analysis would be complicated by the 
heterogeneity of the data and is beyond the scope of this paper but it is clear that, although with a large scatter, NLS1 with 'old' virial BH masses are lying preferentially below the \MBH-\sige\ relation defined by normal galaxies.
When radiation pressure is taken into account in virial BH mass estimates, this tendency disappears or is strongly reduced. It is significant that the NLS1 galaxies with bulge luminosities by \cite{ryan:2007}
are all lying below the expected \MBH-\Lsph\ values while they are in good agreement with it after radiation pressure has been taken into account. 
This is confirmed by Fig.~\ref{fig:histocorr} where we plot the histogram of the distances from the \MBH-\sige\ correlation for the data by \cite{zhou:2006}. In the top left corner we report the mean and standard deviation of the mean ($\sigma/\sqrt{N}$) of residuals from the \MBH-\sige\ correlation. If radiation pressure is not taken into account, NLS1 galaxies lie, on average, a factor $\sim 5$ below the correlation. However, after taking into account radiation pressure, virial BH masses are dispersed around the correlation.

The above findings do not constitute the definitive proof that radiation pressure provides a solution to the small BH mass problem in NLS1. We only show that our calibrated correction for radiation pressure is approximately of the right amount to bring NLS1 to lie on the \MBH-\sige,\Lsph\ correlations.

Finally, although it is not possible to establish whether a system is emitting above Eddington using virial BH masses (see \S\ \ref{sec:physics}), the average increase of BH masses by 0.5-0.7 dex in NLS1 galaxies (from the \citealt{grupe:2004b} and \citealt{zhou:2006} samples, respectively) implies a similar decrease of their classical $L/\LEddO$ ratios.
%%%%%%% FIGURE 5%%%%%%%%%%%%%%%%%%%%%%%%%%%%%%%%%%%%%%%%%%
%%%%%%% FIGURE 5%%%%%%%%%%%%%%%%%%%%%%%%%%%%%%%%%%%%%%%%%%

\section{Summary and Conclusions}\label{sec:summary}

In this paper we have considered the effect of radiation pressure on virial BH mass estimates.

With a simple physical model, we have provided a correction for the effect of radiation pressure on virial products. This correction mainly depends on the average column density \NH\ of broad line clouds.

We have re-calibrated virial BH masses based on reverberation mapping data and single epoch spectra following a procedure analogous to \cite{onken:2004} and \cite{vestergaard:2006}.
With the caveat that it is difficult to accurately quantify the importance of radiation pressure with the present data, we find
consistent values for the radiation pressure correction which, based on the above physical model, indicates an average $\NH\sim\ten{23}\CM\2$ for BLR clouds. This value is remarkably consistent with the BLR cloud column density required in photoionization models to explain the observed spectra.

When taking into account radiation pressure, the average rms scatter of the ratio between single epoch and reverberation mapping virial BH masses drops from  0.4 to 0.2 dex.
The use of single epoch observations as surrogates of expensive reverberation mapping campaigns can thus provide more accurate  virial BH masses than previously thought.

We have considered our newly calibrated virial BH mass  relations for Narrow Line Seyfert 1 galaxies and we have shown that, after taking into account radiation pressure, those galaxies seem to have BH masses similar to that of other broad line AGNs and and which follow the same \MBH-\sige,\Lsph\ relations as normal galaxies.

The small BH masses previously found in NLS1 can be attributed to the neglect 
of radiation pressure in objects radiating close to 
their Eddington limit.

Overall, the analysis presented in this paper clearly indicates that radiation forces arising from the deposition of momentum by ionizing photon constitute an important physical effect which must be taken into account when computing virial BH mass estimates.

\acknowledgments
We thank Julian Krolik for pointing out the importance of radiation pressure from ionizing photons on BLR clouds, 
Hongyan Zhou for providing us with the stellar velocity dispersion measurements of their sample and an anonymous referee for comments and suggestions.
We acknowledge financial contribution from the grant PRIN-MUR 2006025203 by the Italian Ministry of University and Research.

\clearpage


\begin{thebibliography}{68}
\expandafter\ifx\csname natexlab\endcsname\relax\def\natexlab#1{#1}\fi

\bibitem[{{Baldwin} {et~al.}(1995){Baldwin}, {Ferland}, {Korista}, \&
  {Verner}}]{baldwin:1995}
{Baldwin}, J., {Ferland}, G., {Korista}, K., \& {Verner}, D. 1995, \apjl, 455,
  L119+

\bibitem[{{Bentz} {et~al.}(2006{\natexlab{a}}){Bentz}, {Denney}, {Cackett},
  {Dietrich}, {Fogel}, {Ghosh}, {Horne}, {Kuehn}, {Minezaki}, {Onken},
  {Peterson}, {Pogge}, {Pronik}, {Richstone}, {Sergeev}, {Vestergaard},
  {Walker}, \& {Yoshii}}]{bentz:2006}
{Bentz}, M.~C., {Denney}, K.~D., {Cackett}, E.~M., {et~al.} 2006{\natexlab{a}},
  \apj, 651, 775

\bibitem[{{Bentz} {et~al.}(2007){Bentz}, {Denney}, {Cackett}, {Dietrich},
  {Fogel}, {Ghosh}, {Horne}, {Kuehn}, {Minezaki}, {Onken}, {Peterson}, {Pogge},
  {Pronik}, {Richstone}, {Sergeev}, {Vestergaard}, {Walker}, \&
  {Yoshii}}]{bentz:2007}
{Bentz}, M.~C., {Denney}, K.~D., {Cackett}, E.~M., {et~al.} 2007, \apj, 662,
  205

\bibitem[{{Bentz} {et~al.}(2006{\natexlab{b}}){Bentz}, {Peterson}, {Pogge},
  {Vestergaard}, \& {Onken}}]{bentz:2006a}
{Bentz}, M.~C., {Peterson}, B.~M., {Pogge}, R.~W., {Vestergaard}, M., \&
  {Onken}, C.~A. 2006{\natexlab{b}}, \apj, 644, 133

\bibitem[{{Blandford} {et~al.}(1990){Blandford}, {Netzer}, {Woltjer},
  {Courvoisier}, \& {Mayor}}]{blandford:1990}
{Blandford}, R.~D., {Netzer}, H., {Woltjer}, L., {Courvoisier}, T.~J.-L., \&
  {Mayor}, M. 1990, {Active Galactic Nuclei} (Saas-Fee Advanced Course
  20.~Lecture Notes 1990.~Swiss Society for Astrophysics and Astronomy, XII,
  280 pp.~97 figs..~ Springer-Verlag Berlin Heidelberg New York)

\bibitem[{{Botte} {et~al.}(2005){Botte}, {Ciroi}, {di Mille}, {Rafanelli}, \&
  {Romano}}]{botte:2005}
{Botte}, V., {Ciroi}, S., {di Mille}, F., {Rafanelli}, P., \& {Romano}, A.
  2005, \mnras, 356, 789

\bibitem[{{Castor} {et~al.}(1975){Castor}, {Abbott}, \& {Klein}}]{castor:1975}
{Castor}, J.~I., {Abbott}, D.~C., \& {Klein}, R.~I. 1975, \apj, 195, 157

\bibitem[{{Chiang} \& {Murray}(1996)}]{chiang:1996}
{Chiang}, J. \& {Murray}, N. 1996, \apj, 466, 704

\bibitem[{{Collin} \& {Hur{\'e}}(2001)}]{collin:2001}
{Collin}, S. \& {Hur{\'e}}, J.-M. 2001, \aap, 372, 50

\bibitem[{{Collin} \& {Kawaguchi}(2004)}]{collin:2004}
{Collin}, S. \& {Kawaguchi}, T. 2004, \aap, 426, 797

\bibitem[{{Collin} {et~al.}(2006){Collin}, {Kawaguchi}, {Peterson}, \&
  {Vestergaard}}]{collin:2006}
{Collin}, S., {Kawaguchi}, T., {Peterson}, B.~M., \& {Vestergaard}, M. 2006,
  \aap, 456, 75

\bibitem[{{Croton} {et~al.}(2006){Croton}, {Springel}, {White}, {De Lucia},
  {Frenk}, {Gao}, {Jenkins}, {Kauffmann}, {Navarro}, \&
  {Yoshida}}]{croton:2006}
{Croton}, D.~J., {Springel}, V., {White}, S.~D.~M., {et~al.} 2006, \mnras, 365,
  11

\bibitem[{{Dasyra} {et~al.}(2007){Dasyra}, {Tacconi}, {Davies}, {Genzel},
  {Lutz}, {Peterson}, {Veilleux}, {Baker}, {Schweitzer}, \&
  {Sturm}}]{dasyra:2007}
{Dasyra}, K.~M., {Tacconi}, L.~J., {Davies}, R.~I., {et~al.} 2007, \apj, 657,
  102

\bibitem[{{Denney} {et~al.}(2006){Denney}, {Bentz}, {Peterson}, {Pogge},
  {Cackett}, {Dietrich}, {Fogel}, {Ghosh}, {Horne}, {Kuehn}, {Minezaki},
  {Onken}, {Pronik}, {Richstone}, {Sergeev}, {Vestergaard}, {Walker}, \&
  {Yoshii}}]{denney:2006}
{Denney}, K.~D., {Bentz}, M.~C., {Peterson}, B.~M., {et~al.} 2006, \apj, 653,
  152

\bibitem[{{Di Matteo} {et~al.}(2005){Di Matteo}, {Springel}, \&
  {Hernquist}}]{di-matteo:2005}
{Di Matteo}, T., {Springel}, V., \& {Hernquist}, L. 2005, \nat, 433, 604

\bibitem[{Efron \& Tibshirani(1994)}]{efron:1994}
Efron, B. \& Tibshirani, R.~J. 1994, An Introduction to the Bootstrap ({Chapman
  \& Hall/CRC})

\bibitem[{{Elvis}(2000)}]{elvis:2000}
{Elvis}, M. 2000, \apj, 545, 63

\bibitem[{{Everett}(2005)}]{everett:2005}
{Everett}, J.~E. 2005, \apj, 631, 689

\bibitem[{{Ferrarese} \& {Ford}(2005)}]{ferrarese:2005}
{Ferrarese}, L. \& {Ford}, H. 2005, Space Science Reviews, 116, 523

\bibitem[{{Ferrarese} \& {Merritt}(2000)}]{ferrarese:2000}
{Ferrarese}, L. \& {Merritt}, D. 2000, \apjl, 539, L9

\bibitem[{{Gaskell}(1996)}]{gaskell:1996}
{Gaskell}, C.~M. 1996, \apjl, 464, L107+

\bibitem[{{Gebhardt} {et~al.}(2000){Gebhardt}, {Bender}, {Bower}, {Dressler},
  {Faber}, {Filippenko}, {Green}, {Grillmair}, {Ho}, {Kormendy}, {Lauer},
  {Magorrian}, {Pinkney}, {Richstone}, \& {Tremaine}}]{gebhardt:2000}
{Gebhardt}, K., {Bender}, R., {Bower}, G., {et~al.} 2000, \apjl, 539, L13

\bibitem[{{Graham} \& {Driver}(2007)}]{graham:2007a}
{Graham}, A.~W. \& {Driver}, S.~P. 2007, \apj, 655, 77

\bibitem[{{Granato} {et~al.}(2004){Granato}, {De Zotti}, {Silva}, {Bressan}, \&
  {Danese}}]{granato:2004}
{Granato}, G.~L., {De Zotti}, G., {Silva}, L., {Bressan}, A., \& {Danese}, L.
  2004, \apj, 600, 580

\bibitem[{{Greene} \& {Ho}(2005{\natexlab{a}})}]{greene:2005a}
{Greene}, J.~E. \& {Ho}, L.~C. 2005{\natexlab{a}}, \apj, 627, 721

\bibitem[{{Greene} \& {Ho}(2005{\natexlab{b}})}]{greene:2005}
{Greene}, J.~E. \& {Ho}, L.~C. 2005{\natexlab{b}}, \apj, 630, 122

\bibitem[{{Grupe}(2004)}]{grupe:2004a}
{Grupe}, D. 2004, \aj, 127, 1799

\bibitem[{{Grupe} \& {Mathur}(2004)}]{grupe:2004}
{Grupe}, D. \& {Mathur}, S. 2004, \apjl, 606, L41

\bibitem[{{Grupe} {et~al.}(2004){Grupe}, {Wills}, {Leighly}, \&
  {Meusinger}}]{grupe:2004b}
{Grupe}, D., {Wills}, B.~J., {Leighly}, K.~M., \& {Meusinger}, H. 2004, \aj,
  127, 156

\bibitem[{{Jiang} {et~al.}(2007){Jiang}, {Fan}, {Vestergaard}, {Kurk},
  {Walter}, {Kelly}, \& {Strauss}}]{jiang:2007}
{Jiang}, L., {Fan}, X., {Vestergaard}, M., {et~al.} 2007, \aj, 134, 1150

\bibitem[{{Kaspi} {et~al.}(2007){Kaspi}, {Brandt}, {Maoz}, {Netzer},
  {Schneider}, \& {Shemmer}}]{kaspi:2007}
{Kaspi}, S., {Brandt}, W.~N., {Maoz}, D., {et~al.} 2007, \apj, 659, 997

\bibitem[{{Kaspi} {et~al.}(2005){Kaspi}, {Maoz}, {Netzer}, {Peterson},
  {Vestergaard}, \& {Jannuzi}}]{kaspi:2005}
{Kaspi}, S., {Maoz}, D., {Netzer}, H., {et~al.} 2005, \apj, 629, 61

\bibitem[{{Kaspi} \& {Netzer}(1999)}]{kaspi:1999}
{Kaspi}, S. \& {Netzer}, H. 1999, \apj, 524, 71

\bibitem[{{Kaspi} {et~al.}(2000){Kaspi}, {Smith}, {Netzer}, {Maoz}, {Jannuzi},
  \& {Giveon}}]{kaspi:2000}
{Kaspi}, S., {Smith}, P.~S., {Netzer}, H., {et~al.} 2000, \apj, 533, 631

\bibitem[{{Komossa} \& {Xu}(2007)}]{komossa:2007}
{Komossa}, S. \& {Xu}, D. 2007, ArXiv e-prints, 708

\bibitem[{{Korista} \& {Goad}(2004)}]{korista:2004}
{Korista}, K.~T. \& {Goad}, M.~R. 2004, \apj, 606, 749

\bibitem[{{Kormendy} \& {Richstone}(1995)}]{kormendy:1995}
{Kormendy}, J. \& {Richstone}, D. 1995, \araa, 33, 581

\bibitem[{{Krolik}(1999)}]{krolik:1999}
{Krolik}, J.~H. 1999, {Active galactic nuclei : from the central black hole to
  the galactic environment} (Princeton, N.~J.~: Princeton University Press.)

\bibitem[{{Krolik}(2001)}]{krolik:2001}
{Krolik}, J.~H. 2001, \apj, 551, 72

\bibitem[{{Marconi} \& {Hunt}(2003)}]{marconi:2003b}
{Marconi}, A. \& {Hunt}, L.~K. 2003, \apjl, 589, L21

\bibitem[{{Marconi} {et~al.}(2006){Marconi}, {Pastorini}, {Pacini}, {Axon},
  {Capetti}, {Macchetto}, {Koekemoer}, \& {Schreier}}]{marconi:2006}
{Marconi}, A., {Pastorini}, G., {Pacini}, F., {et~al.} 2006, \aap, 448, 921

\bibitem[{{Marconi} {et~al.}(2004){Marconi}, {Risaliti}, {Gilli}, {Hunt},
  {Maiolino}, \& {Salvati}}]{marconi:2004}
{Marconi}, A., {Risaliti}, G., {Gilli}, R., {et~al.} 2004, \mnras, 351, 169

\bibitem[{{Mathews}(1993)}]{mathews:1993}
{Mathews}, W.~G. 1993, \apjl, 412, L17

\bibitem[{{Mathur} \& {Grupe}(2005)}]{mathur:2005}
{Mathur}, S. \& {Grupe}, D. 2005, \aap, 432, 463

\bibitem[{{Mathur} {et~al.}(2001){Mathur}, {Kuraszkiewicz}, \&
  {Czerny}}]{mathur:2001}
{Mathur}, S., {Kuraszkiewicz}, J., \& {Czerny}, B. 2001, New Astronomy, 6, 321

\bibitem[{{McLure} \& {Dunlop}(2002)}]{mclure:2002}
{McLure}, R.~J. \& {Dunlop}, J.~S. 2002, \mnras, 331, 795

\bibitem[{{McLure} \& {Dunlop}(2004)}]{mclure:2004}
{McLure}, R.~J. \& {Dunlop}, J.~S. 2004, \mnras, 352, 1390

\bibitem[{{Murray} \& {Chiang}(1995)}]{murray:1995a}
{Murray}, N. \& {Chiang}, J. 1995, \apjl, 454, L105+

\bibitem[{{Nelson} {et~al.}(2004){Nelson}, {Green}, {Bower}, {Gebhardt}, \&
  {Weistrop}}]{nelson:2004}
{Nelson}, C.~H., {Green}, R.~F., {Bower}, G., {Gebhardt}, K., \& {Weistrop}, D.
  2004, \apj, 615, 652

\bibitem[{{Netzer}(2006)}]{netzer:2006}
{Netzer}, H. 2006, in Lecture Notes in Physics, Berlin Springer Verlag, Vol.
  693, Physics of Active Galactic Nuclei at all Scales, ed. D.~{Alloin}, 1--+

\bibitem[{{Onken} {et~al.}(2004){Onken}, {Ferrarese}, {Merritt}, {Peterson},
  {Pogge}, {Vestergaard}, \& {Wandel}}]{onken:2004}
{Onken}, C.~A., {Ferrarese}, L., {Merritt}, D., {et~al.} 2004, \apj, 615, 645

\bibitem[{{Onken} \& {Peterson}(2002)}]{onken:2002}
{Onken}, C.~A. \& {Peterson}, B.~M. 2002, \apj, 572, 746

\bibitem[{{Peterson}(1997)}]{peterson:1997}
{Peterson}, B.~M. 1997, {An Introduction to Active Galactic Nuclei} (Cambridge,
  New York~: Cambridge University Press)

\bibitem[{{Peterson} \& {Bentz}(2006)}]{peterson:2006}
{Peterson}, B.~M. \& {Bentz}, M.~C. 2006, New Astronomy Review, 50, 796

\bibitem[{{Peterson} {et~al.}(2004){Peterson}, {Ferrarese}, {Gilbert}, {Kaspi},
  {Malkan}, {Maoz}, {Merritt}, {Netzer}, {Onken}, {Pogge}, {Vestergaard}, \&
  {Wandel}}]{peterson:2004}
{Peterson}, B.~M., {Ferrarese}, L., {Gilbert}, K.~M., {et~al.} 2004, \apj, 613,
  682

\bibitem[{{Peterson} \& {Wandel}(2000)}]{peterson:2000}
{Peterson}, B.~M. \& {Wandel}, A. 2000, \apjl, 540, L13

\bibitem[{{Pounds} {et~al.}(1995){Pounds}, {Done}, \& {Osborne}}]{pounds:1995}
{Pounds}, K.~A., {Done}, C., \& {Osborne}, J.~P. 1995, \mnras, 277, L5

\bibitem[{{Proga}(2007)}]{proga:2007b}
{Proga}, D. 2007, \apj, 661, 693

\bibitem[{{Proga} {et~al.}(2000){Proga}, {Stone}, \& {Kallman}}]{proga:2000}
{Proga}, D., {Stone}, J.~M., \& {Kallman}, T.~R. 2000, \apj, 543, 686

\bibitem[{{Ryan} {et~al.}(2007){Ryan}, {De Robertis}, {Virani}, {Laor}, \&
  {Dawson}}]{ryan:2007}
{Ryan}, C.~J., {De Robertis}, M.~M., {Virani}, S., {Laor}, A., \& {Dawson},
  P.~C. 2007, \apj, 654, 799

\bibitem[{{Silk} \& {Rees}(1998)}]{silk:1998}
{Silk}, J. \& {Rees}, M.~J. 1998, \aap, 331, L1

\bibitem[{{Tremaine} {et~al.}(2002){Tremaine}, {Gebhardt}, {Bender}, {Bower},
  {Dressler}, {Faber}, {Filippenko}, {Green}, {Grillmair}, {Ho}, {Kormendy},
  {Lauer}, {Magorrian}, {Pinkney}, \& {Richstone}}]{tremaine:2002}
{Tremaine}, S., {Gebhardt}, K., {Bender}, R., {et~al.} 2002, \apj, 574, 740

\bibitem[{{Tundo} {et~al.}(2007){Tundo}, {Bernardi}, {Hyde}, {Sheth}, \&
  {Pizzella}}]{tundo:2007}
{Tundo}, E., {Bernardi}, M., {Hyde}, J.~B., {Sheth}, R.~K., \& {Pizzella}, A.
  2007, \apj, 663, 53

\bibitem[{{Vestergaard}(2004)}]{vestergaard:2004}
{Vestergaard}, M. 2004, \apj, 601, 676

\bibitem[{{Vestergaard} \& {Peterson}(2006)}]{vestergaard:2006}
{Vestergaard}, M. \& {Peterson}, B.~M. 2006, \apj, 641, 689

\bibitem[{{Willott} {et~al.}(2003){Willott}, {McLure}, \&
  {Jarvis}}]{willott:2003}
{Willott}, C.~J., {McLure}, R.~J., \& {Jarvis}, M.~J. 2003, \apjl, 587, L15

\bibitem[{{Wu} {et~al.}(2004){Wu}, {Wang}, {Kong}, {Liu}, \& {Han}}]{wu:2004}
{Wu}, X.-B., {Wang}, R., {Kong}, M.~Z., {Liu}, F.~K., \& {Han}, J.~L. 2004,
  \aap, 424, 793

\bibitem[{{Zhou} {et~al.}(2006){Zhou}, {Wang}, {Yuan}, {Lu}, {Dong}, {Wang}, \&
  {Lu}}]{zhou:2006}
{Zhou}, H., {Wang}, T., {Yuan}, W., {et~al.} 2006, \apjs, 166, 128

\end{thebibliography}
\end{document}